\documentclass[12ps,epsf]{article}

\usepackage{amsmath}
\usepackage{amssymb}
\usepackage{bm}
\usepackage[dvips]{graphics}

\newcommand{\tr}{{\rm Tr}}

\newcommand{\at}{\tilde{a}}
\newcommand{\vpt}{\tilde{\varphi}}
\newcommand{\bt}{\tilde{b}}
\newcommand{\ct}{\tilde{c}}
\newcommand{\Pt}{{\cal P}}
\newcommand{\Ft}{{\cal F}}

\newcommand{\vpb}{\bar{\varphi}}
\newcommand{\vp}{\varphi}
 
\newcommand{\Slash}[1]{\ooalign{\hfil/\hfil\crcr$#1$}}

\newcommand{\smone}{\sin\left(\frac{\pi m_1}{N_1}\right)} 
\newcommand{\smtwo}{\sin\left(\frac{\pi m_2}{N_1}\right)} 
\newcommand{\smthree}{\sin\left(\frac{\pi m_3}{N_2}\right)} 
\newcommand{\smfour}{\sin\left(\frac{\pi m_4}{N_2}\right)}

\newcommand{\snone}{\sin\left(\frac{\pi n_1}{N_1}\right)} 
\newcommand{\sntwo}{\sin\left(\frac{\pi n_2}{N_1}\right)} 
\newcommand{\snthree}{\sin\left(\frac{\pi n_3}{N_2}\right)} 
\newcommand{\snfour}{\sin\left(\frac{\pi n_4}{N_2}\right)}

\newcommand{\cmnone}{\cos\left(\frac{\pi (m_1+n_1)}{N_1}\right)} 
\newcommand{\cmntwo}{\cos\left(\frac{\pi (m_2+n_2)}{N_1}\right)} 
\newcommand{\cmnthree}{\cos\left(\frac{\pi (m_3+n_3)}{N_2}\right)} 
\newcommand{\cmnfour}{\cos\left(\frac{\pi (m_4+n_4)}{N_2}\right)} 
\date{}

\begin{document}

\begin{titlepage}  \renewcommand{\thefootnote}{\fnsymbol{footnote}}
$\mbox{ }$
\begin{flushright}
\begin{tabular}{l}
KUNS-1953\\
RIKEN-TH-35 \\
hep-th/0412303\\
December 2004
\end{tabular}
\end{flushright}
                                                                                
~~\\
~~\\
~~\\
                                                                                
\vspace*{0cm}
    \begin{Large}
       \vspace{2cm}
       \begin{center}
         {Fuzzy Torus in Matrix Model}
\\
       \end{center}
    \end{Large}
                                                                                
  \vspace{1cm}

\begin{center}
                                                                                
          Subrata B{\sc al}$^a$\footnote
           {
E-mail address : subrata@riken.jp},
          Masanori H{\sc anada}$^{b}$\footnote
           {
E-mail address : hana@gauge.scphys.kyoto-u.ac.jp},
          Hikaru K{\sc awai}$^{ab}$\footnote
           {
E-mail address : hkawai@gauge.scphys.kyoto-u.ac.jp},
          Fukuichiro K{\sc ubo}$^b$\footnote
           {
E-mail address : kubo@gauge.scphys.kyoto-u.ac.jp}

$^a$           {\it Theoretical Physics Laboratory, RIKEN (The
  Institute of Physical and Chemical Research),\\ Wako, Saitama
351-0198, Japan}\\
$^b$           {\it Department of Physics, Kyoto University,
Kyoto 606-8502, Japan}\\
                                                                                
\end{center}

\vfill
                                                                                
\begin{abstract}
\noindent
We have calculated the free energy up to two loop
to compare $T^2$ with $T^4$ in IIB matrix model.
It turns out that $T^2$ has smaller free energy than $T^4$.
 We have also discussed the generation of the gauge group by
 considering $k$-coincident fuzzy tori and found that in this case
 U(1) gauge group is favored. This means that if the true vacuum is
 four-dimensional, it is not a simple fuzzy space considered here.
\end{abstract}
                                                                                
\vfill
\end{titlepage}
\vfil\eject
                                                                                
\setcounter{footnote}{0}

\section{Introduction}
\setcounter{equation}{0}

IIB matrix model \cite{IKK96} is a candidate of the constructive definition of superstring theory. It is defined by the action  
\begin{eqnarray}
  S
  =
  -\frac{1}{g^2}\tr\ 
  \left(
    \frac{1}{4}[A_\mu,A_\nu][A_\mu,A_\nu]
    +
    \frac{1}{2}\bar{\Psi}\Gamma^\mu[A_\mu,\Psi  ]
  \right),
  \label{eq:action}
\end{eqnarray}
where $A_\mu$ and $\Psi$ are $N\times N$ Hermitian matrices. $A_\mu$
is a ten-dimensional vector and $\Psi$ is a ten-dimensional
Majorana-Weyl fermion. 

One of the important aspects of this model is the dynamical
generation of the spacetime and the gauge group. In this model the eigenvalue 
distribution represents spacetime geometry. From this point of view, if a
four-dimensional eigenvalue distribution has smaller free energy
than the other configurations, the spacetime will be compactified to
four-dimensions. On the other hand the stability of $k$ coincident 
D-branes may indicate
the dynamical generation of $U(k)$ gauge group.  
So far, various attempts on this  have been made. 
Several such attempts using  branched polymer \cite{9802085},
complex phase effect \cite{0003223-0108041} and especially
improved mean-field approximation \cite{NS01} were
successful. In these studies, four-dimensional
configurations were found to have smaller free energy. 

Another interesting approach is to compare the free energies on
noncommutative backgrounds. In string theory, noncommutative 
gauge theories on flat or curved backgrounds are realized with 
constant or nonconstant $B_{\mu \nu}$ \cite{stncom}.
IIB matrix model on flat fuzzy space
backgrounds can be mapped to noncommutative super Yang-Mills theory
(NCSYM) \cite{AII99}, and the system can be analyzed by perturbative
calculations. Noncommutative gauge theories on curved manifolds 
have recently been studied extensively by deforming IIB matrix
model \cite{IIBdeform,IIBdeform2}.
 
However, flat fuzzy space backgrounds cannot be
realized as classical solutions at finite $N$. 
Therefore, in order to
see the $N$-dependences of the free energies, we
must study compact noncommutative spaces. Fuzzy $S^2$, $S^2\times S^2$
and $S^2\times S^2\times S^2$ backgrounds have already been studied
and $S^2\times S^2$ has been shown to be stable under
certain deformations \cite{IT03,IKT03}.

In this paper, we study the free energies of IIB matrix model on fuzzy
torus backgrounds. 
The relation to flat fuzzy space is more transparent in the case of
the fuzzy torus than in the case of fuzzy sphere.
We expand the action of IIB matrix
model around a fuzzy torus background and sum up 
all 1PI diagrams. Then, we search the minimum of the free
energy. Especially, we compare $T^2$ and $T^4$ backgrounds. 
In principle, in order
to find the true vacuum, we must calculate free energies on all
possible backgrounds and find the minimum. 
Here, however, we consider only fuzzy torus. 
Although fuzzy torus is not a
classical solution, however, as we will show later, fuzzy torus 
locally looks like flat fuzzy space, which 
is a classical solution. Therefore, we expect our calculation captures
some qualitative feature of flat fuzzy spaces. If the true vacuum is
similar to a flat fuzzy space, we may find some informations 
such as dimensionality of the spacetime on
it. In present study we will see that
$T^2$ has smaller free energy than $T^4$.  This implies that if the
true vacuum is four-dimensional, it is not a  simple noncommutative
space considered here.

Although both fuzzy sphere and fuzzy torus break
supersymmetry, the breaking effect is soft and the loop corrections to
their free energies almost vanish. This is because their local
structures are similar to flat fuzzy spaces. In the case of fuzzy $S^2$,
$S^2\times S^2$ and $S^2\times S^2\times S^2$, such cancellations have
been confirmed to 2-loop level \cite{IT03,IKT03}. 
For fuzzy torus we can see such cancellation at any loop level. 
Furthermore, in this case the calculations become simpler and we can
study larger parameter region. For example, on $T^4$, we vary the
ratio $r$ of the noncommutativity parameters of two tori and their
ratio $R$
of the matrix size \footnote{In the case of $S^2\times S^2$, $r^2=R$
was studied \cite{IT03}. }. We can also see the effect of the global
topologies on the free energies by 
comparing the results for fuzzy tori and fuzzy spheres.

On the other hand, fuzzy torus has a few drawbacks. We embed
fuzzy $T^2$ into $\textbf{R}^4$, while fuzzy $S^2$ can be embedded
into $\textbf{R}^3$. Therefore, in ten dimensions we can realize only
two- and four-dimensional configurations using fuzzy tori ( $T^2$ and
$T^4$), while a
six-dimensional configuration can be constructed using fuzzy spheres (
$S^2\times S^2\times S^2$). In \cite{IKT03} it is shown that
$S^2\times S^2$ has a smaller free energy than $S^2\times S^2\times
S^2$. In our case, such comparison is impossible. We also have
subtlety in the perturbative calculation, due to the presence of
a few tachyons on fuzzy torus.
We will discuss this problem in detail later. 

The organization of the paper is as follows. In section 2, we will 
construct the fuzzy torus backgrounds and show the correspondence
between fuzzy torus and flat fuzzy space. In section 3, we calculate the 
free energy up to two loop level analytically, and in section 4, 
we compare the free energies numerically. In appendices, 
we show the detail of the calculations. Although 
we are interested in $T^4$, we will mostly 
show the Feynman diagrams and calculations of $T^2$ 
as $T^4$ calculations are very lengthy.

\section{IIB matrix model on fuzzy torus  backgrounds}
\setcounter{equation}{0}

In order to perform a perturbative calculation, we decompose
 $A_\mu$ and $\Psi$ into a background $P_\mu$ and 
the fluctuation $a_\mu,\vp$ around it: 
\begin{eqnarray}
A_\mu
=
P_\mu+a_\mu
\quad,\quad
\Psi
=
0+\vp.
\end{eqnarray}
Adding a gauge fixing and  the corresponding ghost term
\begin{eqnarray}
  S_{{\rm GF+FP}}
  =
  -\frac{1}{g^2}
  \tr
  \left(
    \frac{1}{2}[P_\mu,a^\mu]^2+[P_\mu,b][A^\mu,c]
  \right) 
\end{eqnarray}
to (\ref{eq:action}), we obtain the full action 
\begin{eqnarray}
  S+S_{{\rm GF+FP}}
  &=&
  \frac{1}{4g^2}\tr\ F^2
  \nonumber\\
  & &
  +\frac{i}{g^2}\tr\ 
  (\Pt_\mu a_\nu)F_{\mu\nu}
  \nonumber\\
  & &
  +\frac{1}{g^2}\tr\ 
  \left\{
    \frac{1}{2}a_\mu\left(\Pt^2\delta_{\mu\nu}-2i\Ft_{\mu\nu}\right)a_\nu
    -
    \frac{1}{2}\vpb\Slash{\Pt}\vp
    +
    b\Pt^2c
  \right\}\nonumber\\
  & &
  -\frac{1}{g^2}\tr\ 
  \left\{
    (\Pt_\mu a_\nu)[a^\mu,a^\nu]
    +
    \frac{1}{2}\vpb\Gamma^\mu[a_\mu,\vp]
    +
    (\Pt_\mu b)[a^\mu,c]
  \right\}\nonumber\\
  & &
  -\frac{1}{4g^2}\ \tr[a_\mu,a_\nu]^2, 
\end{eqnarray}
where $\Pt_\mu\equiv[P_\mu,\ \cdot\ ],\ F_{\mu\nu}\equiv
i[P_\mu,P_\nu]$ and $\Ft_{\mu\nu}\equiv[F_{\mu\nu},\ \cdot\ ]$. 
If the background is not a classical solution, e.g. fuzzy sphere or
fuzzy torus, the $O(a_\mu)$ term does not vanish.  

\subsection{IIB matrix model on fuzzy $T^2$ background}

First, we introduce the $N\times N$ clock and shift matrices $U$ and $V$
: 
\begin{eqnarray}
  U
  =
  \left(
    \begin{array}{cccc}
      1 & & & \\
      & e^{2\pi i/N} & & \\
      & & e^{2\cdot2\pi i/N} & \\
      & & & \ddots\\
    \end{array}
  \right)
  \ ,\quad
  V
  =
  \left(
    \begin{array}{cccc}
      0 & & &1 \\
      1& 0 & & \\
      &\ddots &\ddots  & \\
      & & 1& 0\\
    \end{array}
  \right)\ , 
\end{eqnarray}
where $U$ and $V$ satisfy
\begin{eqnarray}
  UV
  &=&
  e^{2\pi i/N}VU,
  \quad
  U^nV^m=e^{2\pi i nm/N}V^mU^n,
  \\
  \tr U^n
  &=&
  \tr V^n= 0\quad(n=1,\cdots,N-1),
  \quad
  U^N=V^N=\textbf{1}_N.
\end{eqnarray}

Using them, we construct a 2-dimensional fuzzy torus as follows : 
\begin{eqnarray}
  & &
  P_1+iP_2 = 2\sqrt{N}L_1U
  ,\quad
  P_1-iP_2 = 2\sqrt{N}L_1U^{-1},
  \nonumber\\
  & &
  P_3+iP_4 = 2\sqrt{N}L_2V
  ,\quad
  P_3-iP_4 = 2\sqrt{N}L_2V^{-1},
  \nonumber\\
  & &
  P_1=\sqrt{N}L_1(U+U^{-1})
  ,\quad
  P_2=-i\sqrt{N}L_1(U-U^{-1}),
  \nonumber\\
  & &
  P_3=\sqrt{N}L_2(V+V^{-1})
  ,\quad
  P_4=-i\sqrt{N}L_2(V-V^{-1}),
  \nonumber\\
  & &
  P_5=\cdots=P_{10}=0.
  \label{T2} 
\end{eqnarray}
Here, $L_1$ and $L_2$ are the noncommutativity parameters, which have
the dimension\footnote
  {In IIB matrix model, usually  we treat
  $A_\mu$ as the spacetime coordinate. However, when we map IIB
  matrix model to the non-commutative Yang-Mills theory,
  we usually regard $P_\mu$ as the momentum operator and
  assign the mass dimension to $A_\mu$. }
of mass.  
Their product, $L^2=L_1L_2$, corresponds to the noncommutativity
parameter $B$ of fuzzy plane. 
The relation of this background to fuzzy plane is discussed more precisely
in section \ref{sec:relation_of_torus_to_D_brane}.

We can expand the fluctuations in terms of $U$ and $V$. For example,
$a_\mu$ can be expressed as  
\begin{eqnarray}
  a_\mu
  &=&
  \sum_{m_1,m_2=1}^N\ \at_\mu(m_1,m_2)U^{m_2}V^{-m_1}, 
  \label{eq:Fourier}
\end{eqnarray}
and the hermiticity condition is given by
\begin{eqnarray}
  \at_\mu^\ast(m_1,m_2)
  =
  e^{-2\pi im_1m_2/N}\at_\mu(-m_1,-m_2). 
\end{eqnarray}

We will show that this
expansion corresponds to the ordinary Fourier expansion in section
\ref{sec:relation_of_torus_to_D_brane}.
Using these Fourier modes, we can construct the Feynman rules, 
which we summarize in Appendix \ref{sec:Feynman_rule}.

\subsection{IIB matrix model on fuzzy $T^4$ background}

We can similarly construct fuzzy $T^4$ using 
the tensor products of the fuzzy tori constructed above : 
\begin{eqnarray}
  & &
  P_1=\sqrt{N_1}L_1(U_1+U_1^{-1})\otimes\textbf{1}_{N_2}
  ,\quad
  P_2=-i\sqrt{N_1}L_1(U_1-U_1^{-1})\otimes\textbf{1}_{N_2},
  \nonumber\\
  & &
  P_3=\sqrt{N_1}L_1(V_1+V_1^{-1})\otimes\textbf{1}_{N_2}
  ,\quad
  P_4=-i\sqrt{N_1}L_1(V_1-V_1^{-1})\otimes\textbf{1}_{N_2},
  \nonumber\\
  & &
  P_5=\sqrt{N_2}L_2\ \textbf{1}_{N_1}\otimes(U_2+U_2^{-1})
  ,\quad
  P_6=-i\sqrt{N_2}L_2\ \textbf{1}_{N_1}\otimes(U_2-U_2^{-1}),
  \nonumber\\
  & &
  P_7=\sqrt{N_2}L_2\ \textbf{1}_{N_1}\otimes(V_2+V_2^{-1})
  ,\quad
  P_8=-i\sqrt{N_2}L_2\ \textbf{1}_{N_1}\otimes(V_2-V_2^{-1})\ ,
  \nonumber\\
  & &
  P_9=P_{10}=0. 
\end{eqnarray}
Here, $N=N_1N_2$, and $U_i,V_i$ are the $N_i\times N_i$ clock and shift
matrices. For simplicity, we set the radii of the two circles in the first
$T^2$ to the same value $2\sqrt{N_1}L_1$, and similarly in the second
$T^2$ to $2\sqrt{N_2}L_2$. 
In the following sections, we first fix the value of $N$, and calculate
the free energy as a function 
of $R=\frac{N_2}{N_1}$, $L=\sqrt{L_1L_2}$ and $r=\frac{L_2}{L_1}$, and
minimize it with respect to them.

\subsection{$k$-coincident fuzzy tori}
We also consider a system of $k$-coincident fuzzy tori. To study
$k$-coincident fuzzy $T^2$,  we  replace $U_{(N)}$ and
$V_{(N)}$ with $U_{(N/k)}\otimes\textbf{1}_k$ and
$V_{(N/k)}\otimes\textbf{1}_k$, where $U_{(N)}$ and
$V_{(N)}$ stand for the $N\times N$ clock and shift matrices,
respectively. Then, the Fourier modes $\at_\mu,\vpt,\bt$ and $\ct$ become
$k\times k$ matrices. 

Similarly, in the case of fuzzy $T^4$, we can replace
$U_1\otimes\textbf{1}_{N_2}$ with
$U_1\otimes\textbf{1}_{N_2}\otimes\textbf{1}_k$, and so on, to
construct a $k$-coincident $T^4$. In this case, we have $N=N_1N_2k$.

\subsection{Correspondence between fuzzy torus and flat fuzzy
  space}\label{sec:relation_of_torus_to_D_brane}

In this section, we discuss the relation between the fuzzy torus and
flat fuzzy space \cite{Kim01}. To be specific, we consider the case of fuzzy
$T^2$ and flat fuzzy plane ( flat D1-brane ). Generalization to higher dimensions is
straightforward. 

In order to avoid confusion, we denote fuzzy plane background as $Q_\mu$
and that of fuzzy torus as $P_\mu$. 
The fuzzy plane is characterized by the constant commutators : 
\begin{eqnarray}
  [Q_1,Q_2]=-iB
  ,\quad
  Q_3=\cdots=Q_{10}=0. 
\end{eqnarray}

Let us write $U$ and $V$ as 
\begin{eqnarray}
  U=\exp\left(i\sqrt{\frac{2\pi}{NB}}Q_1^\prime\right),
  \quad
  V=\exp\left(i\sqrt{\frac{2\pi}{NB}}Q_2^\prime\right).
\end{eqnarray} 
Then, if we consider the region where
eigenvalues of $Q_1^\prime$ and $Q_2^\prime$ are close to zero, the condition
$UV=e^{2\pi i/N}VU$ is equivalent to
$[Q_1^\prime,Q_2^\prime]=-iB$. Therefore, $Q_i\simeq Q_i^\prime$ in
this region. On the other hand, if we compare (\ref{T2}) with
\begin{eqnarray}
  Q_1^\prime
  \simeq
  -\frac{i}{2}\sqrt{\frac{NB}{2\pi}}(U-U^{-1})
  ,\quad
  Q_2^\prime
  \simeq
  -\frac{i}{2}\sqrt{\frac{NB}{2\pi}}(V-V^{-1}),
\end{eqnarray}
we have the identifications in the embedded space as $P_2\simeq
const.Q_1^\prime,P_4\simeq const.Q_2^\prime$.
In general, tangent spaces of fuzzy torus look like fuzzy planes.

IIB matrix model on fuzzy torus can be mapped to NCSYM as in the case
of flat fuzzy space \cite{AII99,Kim01}. 
If we introduce the spacetime coordinates as 
\begin{eqnarray}
  \hat{x}_1=-\frac{Q_2}{B}
  ,\quad
  \hat{x}_2=\frac{Q_1}{B},\
\end{eqnarray}
we have
\begin{eqnarray}
  U^{m_2}V^{-m_1}
  &\sim&
  e^{-im_2\sqrt{\frac{2\pi B}{N}}\hat{x}_2}
  e^{-im_1\sqrt{\frac{2\pi B}{N}}\hat{x}_1}
  \nonumber\\
  &=&
  e^{i\sum_{i}k_i\hat{x}_i}\times{\rm (phase)}, 
\end{eqnarray}
where momenta $k_i$ are defined by $k_i=m_i\sqrt{\frac{2\pi B}{N}}$. 
Hence the expansion(\ref{eq:Fourier}) becomes the ordinary Fourier expansion
in the commutative limit.

\section{Calculation of free energy}
\setcounter{equation}{0}

We calculate the free energies in this section. We ignore 
the tachyons and zero modes on the fuzzy torus in our calculations. 
The vector- and fermion-kinetic terms on fuzzy torus have tachyons and extra 
zero-modes besides those coming from the U(1)-part. (See Appendix
\ref{sec:tachyon} for the detail.) 
Therefore, it is not clear whether perturbative calculation is possible or not. 
However, the number of tachyons remains the same in the large-$N$ limit, 
and their mass squared is of $O\left(\frac{1}{N}\right)$ and vanishes in 
the large-$N$ limit. 
Hence we expect that the tachyons give a relative correction of
$O\left(\frac{\log(N)}{N^2}\right)$ to the free energy. 
On the other hand, in the case of $T^4$ and
$N_1\sim N_2\sim \sqrt{N}$, we will see that the leading contributions
cancel out due to the supersymmetry, 
but the terms of $O\left(\frac{1}{N}\right)$ do not vanish. They
are expected to be larger than the effect of 
the tachyons, and we can take the large-$N$ limit without paying much
attention to the tachyons.

\subsection{Tree-level free energy}
We start with the tree-level free energy. It is given by
\begin{eqnarray}
  -\frac{1}{4g^2} \sum_{i,j}
  \tr\ [P_i,P_j]^2
  &=&
  \frac{16L^2N^3}{g^2}\sin^2\left(\frac{\pi}{N}\right)
  \nonumber\\
  &\to&
  \frac{16N\pi^2L^2}{g^2}
  \quad
  \mbox{(in the large-$N$ limit)}
\end{eqnarray}
for $T^2$, and 
\begin{eqnarray}
  -\frac{1}{4g^2} \sum_{i,j}
  \tr\ [P_i,P_j]^2
  &=&
  \frac{16L^2N^3}{g^2}\sin^2\left(\frac{\pi}{N}\right)
  \left(r^2+\frac{1}{r^2}\right)
  \nonumber\\
  &\to&
  \frac{16N\pi^2L^2}{g^2}
  \left(r^2+\frac{1}{r^2}\right)
  \quad
  \mbox{(in the large-$N$ limit)}
  \nonumber\\
\end{eqnarray}
for $T^4$, respectively.

\subsection{1-loop free energy}

The 1-loop free energy can be easily evaluated as \cite{IKK96} 
\begin{eqnarray}
  -\frac{1}{2}\sum_{n=1}^\infty
  \frac{1}{n}\tr\ 
  \left(
    -\frac{2i}{\Pt^2}\Ft_{ij}
  \right)^n
  +
  \frac{1}{4}\sum_{n=1}^\infty
  \frac{1}{n}\tr\ 
  \left(
    \frac{i}{2\Pt^2}\Ft_{ij}\gamma^{ij}
  \right)^n. 
\end{eqnarray}
The terms for $n=1,2,3$ cancel out, and the first nonzero
contribution comes from $n=4$. In the case of $T^4$ 
and $N_1\sim N_2\sim \sqrt{N}$, it is $O(N^0)$ and can be neglected
compared with the tree-level and 2-loop level contributions. In the
case of $T^2$, truncation at some order of $\Ft$ is not a good
approximation because of infrared divergence. 

Instead, we can numerically evaluate the 1-loop free energy without expanding in $\Ft$. 
We calculate all the eigenvalues of the kinetic terms, 
and sum up their logarithms. We find that the 1-loop free energy on
$T^2$ background is proportional to $\log N$ and hence negligible
compared to the tree level and 2-loop contributions.
\footnote{In this calculation, we have discarded the tachyons and the
  extra zero-modes.}

\subsection{2-loop free energy}

Next we evaluate the 2-loop free energy on the fuzzy $T^4$. Since it
is difficult to obtain analytic forms of the 
vector and fermion propagators, we expand them in powers of $\Ft$ as
shown in Appendix \ref{sec:propagators}, 
and calculate the free energy order by order in $\Ft$. A naive power
counting shows that each power of $\Ft$ is associated with
$\frac{1}{N}$ and $\frac{1}{\sqrt{N}}$ for $T^2$ and $T^4$,
respectively, if there is no infrared divergence.
In fact, this expansion is good on the $T^4$ background, while it is
not on the $T^2$ background. 
We discuss the validity of this expansion in Appendix
\ref{sec:F-expansion}. Here we show the result for the $T^4$
background.

\subsubsection{$O(\Ft^0)$}

\begin{figure}[htbp]
  \begin{center}
    \scalebox{0.5}{
      \includegraphics[20cm,7cm]{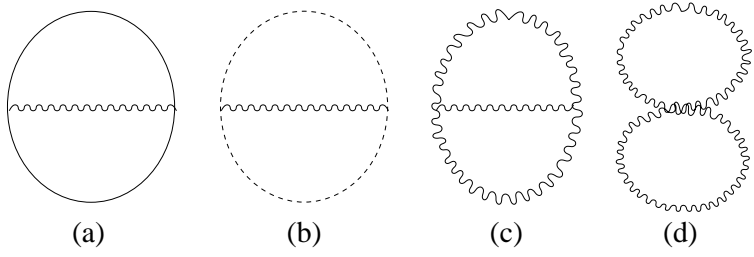}
    }
    \caption{2-loop vacuum diagrams}\label{fig:2loop_diagrams}
  \end{center}
\end{figure}
We have four kinds of 1PI 2-loop vacuum diagrams ( see
Fig.\ref{fig:2loop_diagrams}), 
 and their contributions can be approximated as
(see Appendix \ref{twoener} for the detail) 
\begin{eqnarray}
  {\rm (a)} 
  &\rightarrow &
  \frac{g^2}{8N^2L^4}
  \sum_{m,n}
  \frac{1-\cos(2\pi m\circ n)}{F(m)^2\cdot F(n)^2}, 
  \label{eq:2loop_V3fs}\\
  {\rm (b)} 
  &\rightarrow&
  -\frac{g^2}{512N^2L^4}
  \sum_{m,n}
  \frac{1-\cos(2\pi m\circ n)}{F(m)^2\cdot F(n)^2},
  \label{eq:2loop_V3g_approximateds}\\
  {\rm (c)}
  &\rightarrow&
  \frac{27g^2}{512N^2}
  \sum_{m,n}\frac{1-\cos(2\pi m\circ n)}{F(m)^2\cdot F(n)^2}, 
  \label{eq:2loop_V3b_approximateds}\\
   {\rm (d)}
  &\rightarrow &
  -\frac{45g^2}{256N^2}
  \sum_{m,n}\frac{1-\cos(2\pi m\circ n)}{F(m)^2\cdot F(n)^2}. 
  \label{eq:2loop_V4bs}
\end{eqnarray}
where 
$m\circ n= m_\mu C^{\mu\nu}n_\nu=\frac{m_1 n_2 - m_2 n_1}{N_1} +
\frac{m_3 n_4 - m_4 n_3}{N_2}$, and 
$F(n)^2= \sum_jF_j(n)\cdot F_j(n)$.\\
$F_j(n)$ are analogous to the ordinary momentum variables in the continuum 
limit and given in appendix \ref{twoener} .

Summing up these approximated values (\ref{eq:2loop_V3fs})$\sim$(\ref{eq:2loop_V4bs}), we have
\begin{eqnarray}
  \left(
    \frac{1}{8}
    -
    \frac{1}{512}
    +
    \frac{27}{512}
    -
    \frac{45}{256}
  \right)
  \frac{g^2}{N^2}
  \sum_{m,n}\frac{1-\cos(2\pi m\circ n)}{F(m)^2\cdot F(n)^2}
  =
  0. 
\end{eqnarray}

This cancellation is not exact, but we can see numerically that the exact 
contribution of $O(\Ft^0)$ terms is smaller than that of  $O(\Ft^2)$ terms.

\subsubsection{$O(\Ft^1)$}

The $O(\Ft^1)$ contribution is zero, because integrands are odd function of the momenta. 

\subsubsection{$O(\Ft^2)$}

The calculation of $O(\Ft^2)$ contribution is very tedious. Here we
show only the simplest case, that is, the planar part of
Fig.\ref{fig:2loop_diagrams} (b).
In this case, the contribution can be writen as
\begin{equation}
  \frac{g^2 N}{L^4} f_b(R,r),
\label{of2-form}
\end{equation}
where
\begin{eqnarray}
&&f_b(R,r) \nonumber\\&&  
=  \frac{\pi^2}{256}\int d^4k d^4l
  \biggl\{
  \frac{
    1
    -
    \cos\left(2\pi(k_1+l_1)\right)
    \cos\left(2\pi(k_2+l_2)\right)
  }{F(k)^2\cdot F(l)^2\cdot(F(k+l)^2)^3}
  \nonumber\\
  & &
  \times
  \frac{1}{r^3\sqrt{R}} 
  \left(
    \sin(\pi k_1)\sin(\pi l_1)\cos\left(\pi(k_1+l_1)\right)
    +
    \sin(\pi k_2)\sin(\pi l_2)\cos\left(\pi(k_2+l_2)\right)
  \right)
  \nonumber\\
  & &
  +
  \frac{
      1
      -
      \cos\left(2\pi(k_3+l_3)\right)
      \cos\left(2\pi(k_4+l_4)\right)
    }{F(k)^2\cdot F(l)^2\cdot(F(k+l)^2)^3}
    \nonumber\\
  & &
  \times
  r^3\sqrt{R}
  \left(
    \sin(\pi k_3)\sin(\pi l_3)\cos\left(\pi(k_3+l_3)\right)
    +
    \sin(\pi k_4)\sin(\pi l_4)\cos\left(\pi(k_4+l_4)\right)
  \right) 
  \biggl\}. 
  \nonumber\\
\end{eqnarray}
Here 
$k_i=m_i/N_1,k_j=m_j/N_2,l_i=n_i/N_1,l_j=n_j/N_2\ (i=1,2;j=3,4)$, and  
$k_\mu$ and $l_\mu$ vary from $0$ to $1$. 
Because the ghost propagators do not contain $\Ft$, 
this expression is rather simple. 
The contributions from the other diagrams ( Fig.1 (a),(b),(c) ) also
can be written as 
$\frac{g^2 N}{L^4} f_a(R,r)$,
$\frac{g^2 N}{L^4} f_c(R,r)$ and
$\frac{g^2 N}{L^4} f_d(R,r)$.
But the expressions of $f_a(R,r), f_c(R,r)$ and  $f_d(R,r)$
become much more complicated, because both vector and fermion propagators 
contain 
$O(\Ft)$ corrections.  
Summing up all of the diagrams, we find that
the leading behavior of the $O(\Ft^2)$ contribution can be expressed as 
\begin{eqnarray}
  \frac{g^2N}{L^4}\cdot h(R,r). 
\end{eqnarray}
The function $h(R,r)=\sum_{j=a,b,c,d} f_j(R,r)$ is dimensionless and
has a symmetry under the exchange of two tori
\begin{eqnarray}
  h(R,r)
  =
  h(R^{-1},r^{-1}). 
\end{eqnarray}

\subsection{Total Free energy}

The total free energy on the fuzzy $T^4$ background can be written as 
\begin{eqnarray}
  \frac{16N\pi^2L^4}{g^2}
  \left(
    r^2+\frac{1}{r^2}
  \right)
  +
  \frac{g^2N}{L^4}\cdot h(R,r)
  +
  O(g^4),\  
\end{eqnarray}
where we have neglected subleading terms in $N$. 
The $O(g^4)$ terms come from higher-loop diagrams. If we neglect them and minimize the free energy
with respect to $L$, we obtain the minimum value  
\begin{eqnarray}
  E(R,r;N)
  =
  4\pi N
  \sqrt{\left(r^2+\frac{1}{r^2}\right)h(R,r)} 
\label{totenergy}
\end{eqnarray}
at
\begin{eqnarray}
  \frac{L^4}{g^2}
  =
  \frac{1}{4\pi}
  \sqrt{\frac{h(R,r)}{\left(r^2+\frac{1}{r^2}\right)}}. 
\end{eqnarray}

The situation does not change qualitatively, even if we take the full order of the loop expansion into account.  
In general, the $n$-loop diagrams are proportional to
$\left(\frac{g^2}{L^4}\right)^{n-1}N^2$. However, the $O(\Ft^0)$
contributions cancel due to the supersymmetry\footnote{See Appendix
  \ref{SUSY_cancellation} for the detail. },  and the $O(\Ft^1)$
contributions vanish because they are given by integrals of odd
functions. Therefore, if we include the contribution from the
higher-loop diagrams, the free energy can be expressed as 
\begin{eqnarray}
  N\ 
  \sum_{j=0}^\infty
  \left(\frac{g^2}{L^4}\right)^{j-1}
  h_j(R,r)\ ,
  \label{eq:total_free_energy}
\end{eqnarray}
where $h_j$ are of $O(1)$. The minimum value is again of $O(N)$. 

\subsection{The case of $k$-coincident fuzzy tori}

Here we consider the system of $k$-coincident fuzzy tori. 
This system can be mapped to U($k$) NCSYM. The free energy in this 
case becomes 
\begin{eqnarray}
  kN\ 
  \sum_{j=0}^\infty
  \left(\frac{g^2k}{L^4}\right)^{j-1}
  h_j(R,r). 
\end{eqnarray}
Therefore, the single fuzzy torus has the lowest free energy,  
and $U(1)$ gauge group is favored in this case.

\section{Numerical comparison of free energies}

As the analytic expressions are very complicated, here we try to 
compare the free energies numerically. 
First, we consider $h(R,r)$. It is given by an 8-fold integral, 
and we have evaluated it approximately using the quadrature by parts. 

We divide the integration domain into $M$ bins in each direction, with $M=9,10,\cdots,15$. 
Let $h(R,r;M)$ be the value of $h(R,r)$ evaluated by $M$ bins. 
We have fitted the plot of $h(R,r;M)$ versus $1/M$ by a polynomial, 
and estimated $h(R,r)=h(R,r;\infty)$ extrapolating it up to $1/M \to 0$.
\begin{figure}[htbp]
  \begin{center}
    \scalebox{0.7}{
      \includegraphics[15cm,10cm]{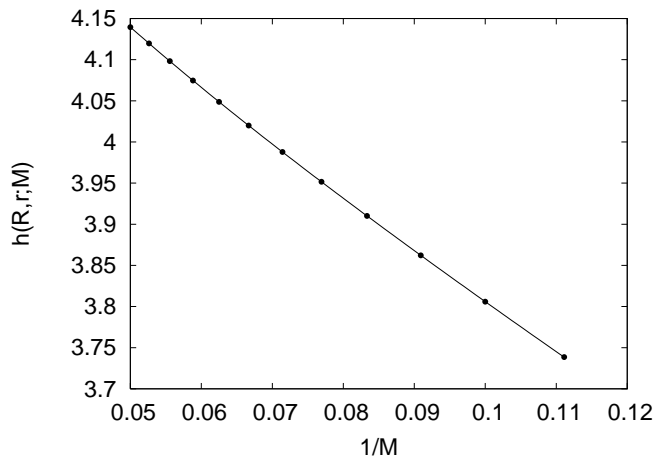}
    }
    \caption{$h(R,r;M)$ versus $1/M$ for $R=r=1$. The curve shows a nice
    polynomial fitting.}
    \label{fig:fitting}
  \end{center}
\end{figure}
In Fig.\ref{fig:fitting}, we plot $h(R,r;M)$ against  $1/M$ for 
$R=1$ and $r=1$. 
The curve shows a nice polynomial fitting. We can evaluate the 
$h(R,r)=h(R,r;\infty)$ replacing $1/M = 0$ in the polynomial.

Once we have $h(R,r)$, we can readily obtain $E(R,r;N)$ from 
eqn. (\ref{totenergy}). 
\begin{figure}[htbp]
  \begin{center}
    \scalebox{0.7}{
      \includegraphics[15cm,10cm]{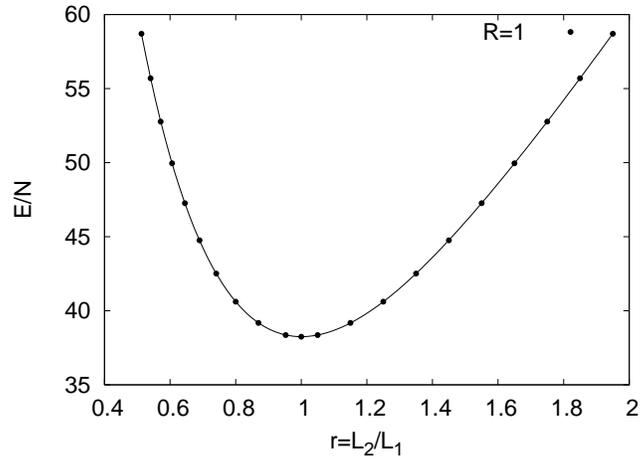}
    }
    \caption{$E(R,r;N)/N$  against $r$ for fixed $R=1$. 
     $R=r=1$ is the minimum here. $T^4$ is favored under variation of $r$.}\label{fig:R=1}
  \end{center}
\end{figure}
We plot $E(R=1,r;N)/N$ as a function of $r$ for $R$ fixed to 1 in
Fig.\ref{fig:R=1}. 
We see that $R=r=1$ is the minimum. 
Therefore, the fuzzy $T^4$ is favored under the variation of the ratio 
$r$ of the noncommutativity.

\begin{figure}[htbp]
  \begin{center}
    \scalebox{0.7}{
      \includegraphics[15cm,10cm]{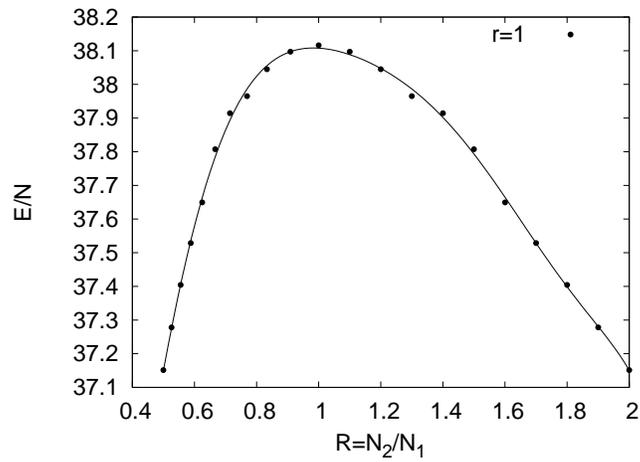}
    }
    \caption{$E(R,r;N)/N$ , against $R$ for fixed $r=1$.  $R=r=1$
     is the maximum here. $T^4$ unstable under variation of $R$.}\label{fig:r=1}
  \end{center}
\end{figure}

Fig.\ref{fig:r=1} shows $E(R,r=1;N)/N$ as a function of $R$ for
$r$ fixed to 1, and this time $R=r=1$ is the maximum. 
Therefore, we find that the fuzzy $T^4$ is {\itshape unstable} under
the variation of the ratio $R$ of the matrix sizes. 

\begin{figure}[htbp]
  \begin{center}
    \scalebox{0.7}{
      \includegraphics[15cm,10cm]{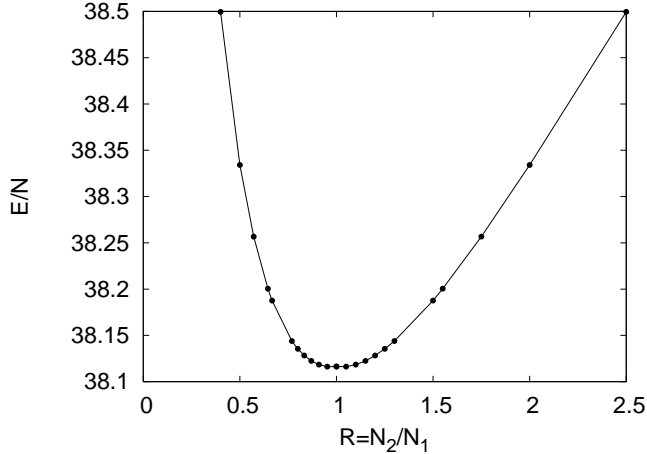}
    }
    \caption{$E(R,r;N)/N$  against $R$ for $R=r^2$. 
     $R=r=1$ is the minimum here. $T^4$ is favored.}\label{fig:R=r^2}
  \end{center}
\end{figure}

To compare with the result for the fuzzy $S^2\times S^2$ in \cite{IT03}, we plot 
$E(R,r=\sqrt{R};N)/N$ as a function of $r$ in Fig.\ref{fig:R=r^2}.
The curve $r=\sqrt{R}$ is the same as the one studied for the 
fuzzy $S^2\times S^2$ in \cite{IT03}. 
On this curve, $R=r=1$ is the minimum, hence the fuzzy $T^4$ is
favored. This observation is similar to the fuzzy $S^2\times S^2$. 
In \cite{IT03}, it is pointed out that 
although the 2-loop contribution ($h(R,r=\sqrt{R})$ in our notation) is maximum 
at $R=r=1$, the tree-level contribution, $\sim
r^2+\frac{1}{r^2}=R+\frac{1}{R}$,  
is minimum at $R=r=1$ and this makes the total free energy minimum at $R=r=1$. 
We have exactly the same situation here.

These results show that although the fuzzy $T^4$ is favourable in
some directions, it is unstable under the variation of $R$. Because the limit 
$R\to 0\ {\rm or}\ \infty$ corresponds to the fuzzy $T^2$, we conclude that 
the fuzzy $T^2$ is more favourable than the fuzzy $T^4$. Since our 
observations along $r=\sqrt{R}$ is completely analogous to that of 
\cite{IT03}, it is plausible that the fuzzy $S^2\times S^2$ also has 
the same kind of instability. 
A similar phenomenon is observed  in a 6 dimensional bosonic matrix
model with Chern Simons term, where 
$S^2$ is indeed shown to be more favourable than $S^2 \times S^2$
\cite{azubals2s2}.

We can compare the free energy of the fuzzy $T^4$
and the fuzzy $S^2\times S^2$. At $R=r=1$, the free energy of $T^4$
is $\simeq 38.1N$. According to \cite{IT03}, the free energy of
$S^2\times S^2$ with the same matrix size and noncommutativity is
$\simeq 3.6N$. Therefore, $S^2\times S^2$ is more favorable than
$T^4$. 

We have also computed the 2-loop free energy of the fuzzy $T^2$ up to
$O(\Ft^2)$ as a function of the ratio of radii, $L_1/L_2$. We found
that it has the minimum at $L_1=L_2$ and is proportional to
$N$. However, because the $\Ft$ expansion is not good in this
background, this result is not reliable. The free energy on the $S^2$
background is calculated in \cite{IKT03}. In this case, the tree-level
action dominates and the free energy is $0$ at $L=0$.

\section{Conclusions and discussions}

In this paper, we have calculated the free energies of IIB matrix
model on the fuzzy $T^4$ backgrounds up to 2-loop level. The free
energy on $T^4$ is a function of the ratio $r$ of the noncommutativity
and $R$ of the matrix size. At $R=r=1$, the free energy takes minimum
when $r$ is varied, but maximum when $R$ is varied. This
indicates that $T^2$ is more stable than $T^4$. It is plausible that
$S^2\times S^2$ has the same sort of instability. If so, a
four-dimensional vacuum might not be realized as a direct product of
two noncommutative surfaces. Commutative backgrounds seem to be more
plausible. Another possibility is that new terms like Myers term arise
dynamically (for example via large-$N$ RG ) and stabilize
noncommutative spaces. 

We also have compared the free energies of $S^2\times S^2$ and $T^4$, and
found that the latter is about 10 times as large as the former. It
implies that IIB matrix model favors configurations with higher
symmetry.  

Furthermore, we calculated the free energy of the system of
$k$-coincident fuzzy $T^4$, and found that the $k=1$ system has the
smallest free energy. In terms of NCSYM, this means that U(1) gauge
group is favored. This result is the same as that of the fuzzy sphere
in IIB matrix model \cite{IT03} and in bosonic matrix models 
with Chern Simons term \cite{azubalnn}.
It seems that in general noncommutative backgrounds favor
U(1) gauge group.

\begin{center} \begin{large}
Acknowledgments
\end{large} \end{center}
The authors would like to thank Y.Kimura, Y.Matsuo and A.Miwa for
fruitful discussions.
S.B. was supported in part by the Grants-in-Aid for 
Scientific Research (No. P02040 )  the Ministry of Education, 
Culture, Sports, Science and Technology of Japan. M.H. would like to
thank the Japan Society for the Promotion of Science for financial
support. This work was also supported in part by a Grant-in-Aid for
the 21st Century COE ``Center for Diversity and Universality in
Physics''.

\appendix

\section{Feynman rules of IIB matrix model on fuzzy $T^2$ background}\label{sec:Feynman_rule}
\setcounter{equation}{0}

In this section, we summarize the Feynman rules of IIB matrix model on
the $T^2$ background. Generalization to the $T^4$ background is
straightforward.   

\subsection{Propagators}\label{sec:propagators}

\textbf{Boson ( vector and scalar )}\\ \\
The kinetic term for $a_{\mu}$ is given by  
\begin{eqnarray}
  \lefteqn{
  \frac{1}{2g^2}\tr\ 
  a_\mu
  \left\{
    \Pt^2\delta_{\mu\nu}
    -
    2i\Ft_{\mu\nu}
  \right\}
  a_\nu
  }\nonumber\\
  &=&
  \frac{N}{2g^2}\sum_{m,n}
  e^{-2\pi im_1m_2/N}
  \at_\mu(-m_1,-m_2)
  \left\{
    \Pt^2\delta_{\mu\nu}
    -
    2i\Ft_{\mu\nu}
  \right\}_{m_1m_2;n_1n_2}
  \at_\nu(n_1,n_2)\ ,
  \nonumber\\
\end{eqnarray}
and the propagator is expressed as
\begin{eqnarray}
  \langle\at_\mu(-m_1,-m_2)\at_\nu(n_1,n_2)\rangle
  =
  \frac{g^2}{N}
  \left\{
    \Pt^2\delta_{\mu\nu}
    -
    2i\Ft_{\mu\nu}
  \right\}^{-1}_{m_1m_2;n_1n_2}
  \times
  e^{2\pi in_1n_2/N}. 
\end{eqnarray}
We expand it with respect to $\Ft$ and truncate it at some order : 
\begin{eqnarray}
  \lefteqn{
    \left\{
      \Pt^2\delta_{\mu\nu}
      -
      2i\Ft_{\mu\nu}
    \right\}^{-1}
  }\nonumber\\
  &=&
  \left\{
    \frac{\delta_{\mu\nu}}{\Pt^2}
    +
    2i\frac{1}{\Pt^2}\Ft_{\mu\nu}\frac{1}{\Pt^2}
    -
    4\frac{1}{\Pt^2}\Ft_{\mu\rho}
    \frac{1}{\Pt^2}\Ft_{\rho\nu}
    \frac{1}{\Pt^2}
    +\cdots
  \right\}. 
\end{eqnarray}
Note that this expansion is not good in this case but a similar expansion in 
the $T^4$ background is reliable. 
\\ \\ \textbf{Fermion} \\ \\ 
The fermion kinetic term is given by
\begin{eqnarray}
  -\frac{1}{2g^2}\tr\ 
  \vpb\Slash{\Pt}\vp
  =
  -\frac{N}{2g^2}\sum_{m,n}\ 
  e^{-2\pi im_1m_2/N}
  \vp^{{\rm T}}(-m_1,-m_2)\Slash{\Pt}_{m_1m_2;n_1n_2}\vp(n_1,n_2),
\end{eqnarray}
and the propagator is expressed as
\begin{eqnarray}
  \langle\vpt_i(-m_1,-m_2)\vpt_j(n_1,n_2)\rangle
  =
  \frac{g^2}{N}\left(\frac{1}{\Slash{\Pt}}\right)_{ij;m_1m_2;n_1n_2}
  \times
  e^{2\pi in_1n_2}. 
\end{eqnarray}
Here we also expand $\frac{1}{\Slash{\Pt}}$ in powers of $\Ft$ : 
\begin{eqnarray}
  \frac{1}{\Slash{\Pt}}
  &=&
  \left\{
    \Pt^2
    +
    \frac{i}{2}\gamma^{\mu\nu}\Ft_{\mu\nu}
  \right\}^{-1}
  \Slash{\Pt}
  \nonumber\\
  &=&
  \left\{
    \frac{1}{\Pt^2}
    -
    \frac{i}{2}\frac{1}{\Pt^2}\gamma^{\mu\nu}\Ft_{\mu\nu}\frac{1}{\Pt^2}
    -
    \frac{1}{4}\frac{1}{\Pt^2}\gamma^{\mu\nu}\Ft_{\mu\nu}
    \frac{1}{\Pt^2}\gamma^{\rho\sigma}\Ft_{\rho\sigma}
    \frac{1}{\Pt^2}
    +\cdots
  \right\}
  \Slash{\Pt}. 
  \nonumber\\
\end{eqnarray}
\\
\\
\textbf{Ghost}
\\
The Ghost kinetic term and propagator are given, respectively, by 
\begin{eqnarray}
  \frac{1}{g^2}\tr\ 
  b\Pt^2c
  =
  \frac{N}{g^2}\sum_{m,n}
  e^{-2\pi im_1m_2/N}
  \bt(-m_1,-m_2)
  \left(\Pt^2\right)_{m_1m_2;n_1n_2}
  \ct(n_1,n_2),
  \nonumber\\
\end{eqnarray}
and

\begin{eqnarray}
  \langle\bt(-m_1,-m_2)\ct(n_1,n_2)\rangle
  =
  \frac{g^2}{N}
  \left(\Pt^2\right)^{-1}_{m_1m_2;n_1n_2}
  \times
  e^{2\pi in_1n_2/N}. 
\end{eqnarray}

\subsection{Explicit form of $\Pt$}
A direct calculation shows that 
\begin{eqnarray}
    [P_1,a_\mu]
  &=&
  -2i\sqrt{N}L_1\sum_{m}U^{m_2}V^{-m_1}\sin\left(\frac{\pi m_1}{N}\right)
  \left\{
    \at_\mu(m_1,m_2-1)e^{\pi i m_1/N}
    -
    \at_\mu(m_1,m_2+1)e^{-\pi i m_1/N}
  \right\},
  \nonumber\\
  \ [P_2,a_\mu]
  &=&
  -2\sqrt{N}L_1\sum_{m}U^{m_2}V^{-m_1}\sin\left(\frac{\pi m_1}{N}\right)
  \left\{
    \at_\mu(m_1,m_2-1)e^{\pi i m_1/N}
    +
    \at_\mu(m_1,m_2+1)e^{-\pi i m_1/N}
  \right\},
  \nonumber\\
  \ [P_3,a_\mu]
  &=&
  -2i\sqrt{N}L_2\sum_{m}U^{m_2}V^{-m_1}\sin\left(\frac{\pi m_2}{N}\right)
  \left\{
    \at_\mu(m_1+1,m_2)e^{-\pi i m_2/N}
    -
    \at_\mu(m_1-1,m_2)e^{\pi i m_2/N}
  \right\},
  \nonumber\\
  \ [P_4,a_\mu]
  &=&
  -2\sqrt{N}L_2\sum_{m}U^{m_2}V^{-m_1}\sin\left(\frac{\pi m_2}{N}\right)
  \left\{
    \at_\mu(m_1+1,m_2)e^{-\pi i m_2/N}
    +
    \at_\mu(m_1-1,m_2)e^{\pi i m_2/N}
  \right\}. 
  \nonumber\\
\end{eqnarray}
From these equations, we obtain 
\begin{eqnarray}
  \left(\Pt_1\right)_{m_1m_2;n_1n_2}
  =
  \left\{
    \begin{array}{cc}
      \pm 2i\sqrt{N}L_1\sin\left(\frac{\pi m_1}{N}\right)
      e^{\mp\pi i m_1/N}
      & ((n_1,n_2)=(m_1,m_2\pm 1))\\
      0&({\rm otherwise})
    \end{array}
  \right. ,
\end{eqnarray}
and so on. This expression is not very simple, but $\Pt^2$ can be
written in the following simple form : 
\begin{eqnarray}
  \left(\Pt^2\right)_{m_1m_2;n_1n_2}
  =
  16N\sum_iL_i^2
  \sin^2\left(\frac{\pi m_i}{N}\right)
  \delta_{m_1n_1}
  \delta_{m_2n_2}. 
\end{eqnarray}

\subsection{Interaction vertices}

\begin{figure}[htbp]
  \begin{center}
    \scalebox{0.7}{
      \includegraphics[25cm,5cm]{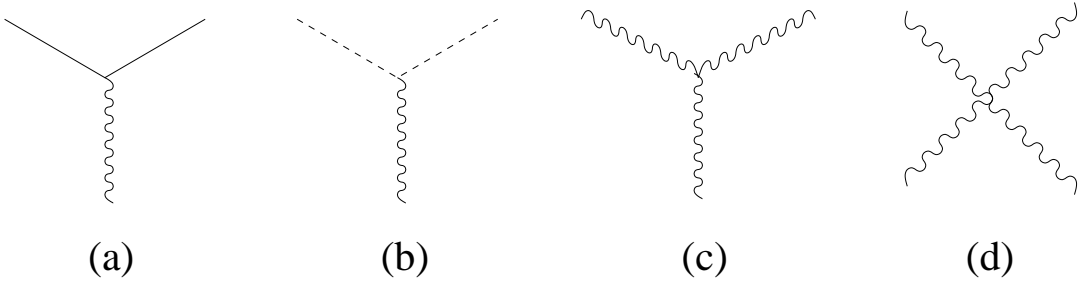}
    }
    \caption{Interaction vertices}\label{fig:vertices}
  \end{center}
\end{figure}

There are four types of interaction vertices ( Fig.\ref{fig:vertices} ). 
Here, the solid, dotted and wavy lines represent the fermions, ghosts
and bosons, respectively.    
\\
\\
\textbf{(a) Fermion-fermion-boson coupling $V_{3f}$}

\begin{eqnarray}
  V_{3f}
  &=&
  -\frac{iN}{g^2}\sum_{k,l}
  \vpt_i(k)(\gamma^\mu)_{ij}\at_\mu(-k-l)\vpt_j(l)
  e^{-\pi i\{k_1k_2+l_1l_2+(k_1+l_1)(k_2+l_2)\}/N}
  \sin(\pi k\circ l),
  \nonumber\\
  C^{ij}
  &\equiv&
  \frac{1}{N}
  \left(
    \begin{array}{cc}
      0 & 1\\
      -1 & 0
    \end{array}
  \right)_{ij}
  ,\quad
  k\circ l
  =
  \sum_{i,j}k_iC^{ij}l_j
  =
  \frac{k_1l_2-k_2l_1}{N}. 
\end{eqnarray}
Note that $\mu=1,2,\cdots,10$
 ( Both vectors and scalars couple to the fermion ). 
\\
\\
\textbf{(b) Ghost-ghost-vector coupling $V_{3g}$}

\begin{eqnarray}
  V_{3g}
  &=&
  2i\frac{(2N)^{3/2}}{g^2}
  \sum_{m,n}
  e^{-\pi i\left\{m_1m_2+n_1n_2+(m_1+n_1)(m_2+n_2)\right\}/N}
  \bt(m)\ct(n)
  \nonumber\\
  & &
  \times
  \biggl\{
  L_1\sin\left(\frac{\pi m_1}{N}\right)
  e^{-\pi i (m_1+n_1)/N}
  \sin\left(\pi m\circ n - \frac{\pi n_1}{N}\right)
  \at_{1+}(-m_1-n_1,-m_2-n_2-1)
  \nonumber\\
  & &
  \quad
  +
  L_1\sin\left(\frac{\pi m_1}{N}\right)
  e^{\pi i (m_1+n_1)/N}
  \sin\left(\pi m\circ n + \frac{\pi n_1}{N}\right)
  \at_{1-}(-m_1-n_1,-m_2-n_2+1)
  \nonumber\\
  & &
  \quad
  -
  L_2\sin\left(\frac{\pi m_2}{N}\right)
  e^{-\pi i (m_2+n_2)/N} 
  \sin\left(\pi m\circ n + \frac{\pi n_2}{N}\right)
  \at_{2-}(-m_1-n_1-1,-m_2-n_2)
  \nonumber\\
  & &
  \quad
  -
  L_2\sin\left(\frac{\pi m_2}{N}\right)
  e^{\pi i (m_2+n_2)/N}
  \sin\left(\pi m\circ n - \frac{\pi n_2}{N}\right)
  \at_{2+}(-m_1-n_1+1,-m_2-n_2)
  \biggl\},
  \nonumber\\
\end{eqnarray}
where
\begin{eqnarray}
  \at_{1\pm}
  \equiv
  \frac{\pm i\at_1+\at_2}{\sqrt{2}}
  ,\quad
  \at_{2\pm}
  \equiv
  \frac{\mp i\at_3+\at_4}{\sqrt{2}}. 
\end{eqnarray}
Note that only vectors couple to the ghosts and scalars do not. 
\\
\\
\textbf{(c) Boson-boson-vector coupling $V_{3b}$}

\begin{eqnarray}
  V_{3b}
  &=&
  2i\frac{(2N)^{3/2}}{g^2}
  \sum_{m,n}
  e^{-\pi i\left\{m_1m_2+n_1n_2+(m_1+n_1)(m_2+n_2)\right\}/N}
  \at_\mu(m)\at^\mu(n)
  \nonumber\\
  & &
  \times
  \biggl\{
  L_1\sin\left(\frac{\pi m_1}{N}\right)
  e^{-\pi i (m_1+n_1)/N}
  \sin
  \left(
    \pi m\circ n-\pi n_1/N
  \right) 
  \at_{1+}(-m_1-n_1,-m_2-n_2-1)
  \nonumber\\
  & &
  \quad
  +
  L_1\sin\left(\frac{\pi m_1}{N}\right)
  e^{\pi i (m_1+n_1)/N}
  \sin
  \left(
    \pi m\circ n+\pi n_1/N
  \right)
  \at_{1-}(-m_1-n_1,-m_2-n_2+1)
  \nonumber\\
  & &
  \quad
  -
  L_2\sin\left(\frac{\pi m_2}{N}\right)
  e^{-\pi i (m_2+n_2)/N} 
  \sin
  \left(
    \pi m\circ n+\pi n_2/N
  \right)
  \at_{2-}(-m_1-n_1-1,-m_2-n_2)
  \nonumber\\
  & &
  \quad
  -
  L_2\sin\left(\frac{\pi m_2}{N}\right)
  e^{\pi i (m_2+n_2)/N}
  \sin
  \left(
    \pi m\circ n-\pi n_2/N
  \right)
  \at_{2+}(-m_1-n_1+1,-m_2-n_2)
  \biggl\}. 
  \nonumber\\
\end{eqnarray}
\\
\textbf{(d) four-boson coupling $V_{4b}$}

\begin{eqnarray}
  V_{4b}
  &=&
  \frac{N}{4g^2} 
  \sum_{k,l,m,n} 
  \at_\mu(k)\at_\nu(l)\at_\mu(m)\at_\nu(n)\delta^2_{k+l+m+n,0}
  e^{-\pi i(k_1k_2+l_1l_2+m_1m_2+n_1n_2)}
  \nonumber\\
  & &
  \times
  \biggl\{
  e^{\frac{\pi i}{2}(k\circ l+l\circ m+m\circ n+n\circ k)}
  +
  e^{-\frac{\pi i}{2}(k\circ l+l\circ m+m\circ n+n\circ k)}
  \nonumber\\
  & &
  \hspace{0.5cm}
  -
  e^{\frac{\pi i}{2}(k\circ l+l\circ n+n\circ m+m\circ k)}
  -
  e^{-\frac{\pi i}{2}(k\circ l+l\circ n+n\circ m+m\circ k)}
  \biggl\}. 
\end{eqnarray}

\section{Cancellation of free energy on torus background due to SUSY
  at $O(\Ft^0)$}\label{SUSY_cancellation}
\setcounter{equation}{0}

First, we consider the cancellation of the free energy on flat fuzzy
space backgrounds. On this background, the free energy should be zero
at least formally. ( The meaning of the word ``formally" is explained
shortly. ) 

For example, the sum of the two-loop diagrams is proportional to \cite{IKT03}
\begin{eqnarray}
  \int dk\ dl
  \left\{
    \frac{1}{k^2\cdot l^2}
    +
    \frac{2k\cdot l}{k^2\cdot l^2\cdot (k+l)^2}
  \right\}.  
  \label{eq:2loop_D_brane}
\end{eqnarray}
Naively this is zero, because $2k\cdot l = (k+l)^2-k^2-l^2$. However,
if the size $N$ of the matrices is finite, there should be some
momentum cutoff, and (\ref{eq:2loop_D_brane}) is not necessarily
zero. For example, in the region where $k$ and $l$ are allowed but
$k+l$ is outside the cutoff, the second term of
(\ref{eq:2loop_D_brane}) does not make sense. However, such problem
occurs only in the large-momentum region. On the other hand in the
small-momentum region, the free energy vanishes due to the
supersymmetry. Therefore, it is natural to expect that the free energy
is ``formally" zero at any order of the loop expansion, if we use
$2k\cdot l = (k+l)^2-k^2-l^2$ and shift the integration momentum not
taking into account the cutoff. We assume this property.

Next, we consider the fuzzy torus background. Note that in this case
the cutoff procedure is completely determined and there is no
ambiguity. 
To be concrete, we consider the $T^2$-background here. If we use 
the approximations such as $\at_\mu(m_1,m_2\pm 1)\simeq\at_\mu(m_1,m_2)$,
we have 
\begin{eqnarray}
  \Pt_i\at_\mu(m)
  =
  4\sqrt{N}f_i(m)\at_\mu(m),
\end{eqnarray}
where
\begin{eqnarray}
  f_1(m)
  &\equiv&
  L_1\sin^2\left(\frac{\pi m_1}{N}\right),
  \nonumber\\
  f_2(m)
  &\equiv&
  -
  L_1\sin\left(\frac{\pi m_1}{N}\right)
  \cos\left(\frac{\pi m_1}{N}\right),
  \nonumber\\
  f_3(m)
  &\equiv&
  -L_2\sin^2\left(\frac{\pi m_2}{N}\right),
  \nonumber\\
  f_4(m)
  &\equiv&  
  -L_2\sin\left(\frac{\pi m_2}{N}\right)
  \cos\left(\frac{\pi m_2}{N}\right). 
\end{eqnarray}
\\
Under this approximation, at $O(\Ft^0)$, the only difference from
fuzzy plane is that the momentum $\frac{\pi m_i}{N}$ is replaced with
$f_j(m)$. 
\\
The $f_j(m)$'s satisfy the following set of equations : 
\begin{eqnarray}
  f^2(m)
  &=&
  \sum_iL_i^2\sin^2\left(\frac{\pi m_i}{N}\right),
  \\
  2f(m)\cdot f(-n)
  &=&
  f^2(m)+f^2(n)-f^2(m+n),
  \label{eq:inner_product}
\end{eqnarray}
which is an analogue of $-2k\cdot l =k^2+l^2-(k+l)^2$. Using this, we
can express the free energy only by $f^2(m)$'s.  Since the momentum
space is completely periodic 
in the fuzzy torus background, we can shift the integration variables without
any obstruction, and find that the free energy is zero at this level.
If we evaluate the free energy without using the approximations, there
appear $O\left(\frac{1}{N}\right)$-corrections.

\section{Validity of $\Ft$-expansion}\label{sec:F-expansion}
\setcounter{equation}{0}

In this section, we discuss the validity of the $\Ft$-expansion. First
we consider the case of $T^4$. $\Ft$'s are expressed in the momentum
basis such as 
\begin{eqnarray}
  \lefteqn{
    \left(\Ft_{1+,2+}\right)_{m_1-1,m_2-1,m_3,m_4;m_1,m_2,m_3,m_4}
  }\nonumber\\
  &=&
  -8i\sin\left(\frac{\pi}{N_1}\right)
  e^{\pi i/N_1}N_1L_1^2e^{-\pi i(m_1+m_2)/N_1}
  \sin\left(\frac{\pi(m_1-m_2)}{N_1}\right)
  \nonumber\\
  &\sim&
  \sin\left(\frac{\pi(m_1-m_2)}{N_1}\right), 
\end{eqnarray}
and $\Pt^2$ is expressed as 
\begin{eqnarray}
  \lefteqn{
    \left(\Pt^2\right)_{m_1,m_2,m_3,m_4;n_1,n_2,n_3,n_4}
  }\nonumber\\
  &=&
  16
  \left\{
    \sum_{i=1,2}N_1L_1^2
    \sin^2\left(\frac{\pi m_i}{N_1}\right)
    +
    \sum_{i=3,4}N_2L_2^2
    \sin^2\left(\frac{\pi m_i}{N_2}\right)
  \right\}\delta^4_{mn}. 
  \nonumber\\
\end{eqnarray}
Hence, in the generic region, where $\frac{m_1}{N_1}=O(1)$
etc, one $\frac{1}{\Pt^2}\Ft$-insertion amounts to multiplying $\frac{1}{N_1+N_2}\lesssim\frac{1}{\sqrt{N}}$.

However, if the arguments in $\sin$ are small, the situation
changes. For example, if we take $N_1\sim N_2\sim\sqrt{N}$ and
consider the region where all arguments in $\sin$ are of $O(N^{-\alpha})\
(0<\alpha\le\frac{1}{2})$, then $\frac{1}{\Pt^2}\Ft\sim
N^{\alpha-1/2}$. This shows that in the IR region convergence of the
$\Ft$-expansion becomes worse. 
In the case that $N_1\sim N_2\sim\sqrt{N}$, the free energy does not
have IR divergence. Therefore the generic region dominates, and the
convergence of this expansion is good. Indeed, from the numerical
calculations of the 2-loop free energy, we can see that one insertion
of $\frac{1}{\Pt^2}\Ft$ decreases 
the power of $N$ by $\frac{1}{2}$. 
On the other hand, in the case of $T^2$ background, the free energy
has IR divergence. Therefore IR region dominates and the insertion of
$\frac{1}{\Pt^2}\Ft$ does not decrease the power of $N$. In this case
$\Ft$-expansion is not reliable.

\section{Tree-level tachyons and extra zero modes}\label{sec:tachyon}
\setcounter{equation}{0}

On fuzzy torus, there exist zero and tachyonic modes, which we list
below. 

\subsection{ fermion zero modes}

To be concrete we discuss the $T^2$-case here. We use the following 
4-dimensional gamma matrices:
\begin{eqnarray}
  \gamma^\mu
  &=&
  i
  \left(
    \begin{array}{cc}
      0 & -\sigma^{\mu\dagger} \\
      \sigma^\mu & 0
    \end{array}
  \right),
\end{eqnarray}
where $\sigma^i(i=1,2,3)$ are the Pauli matrices and $\sigma^4=i\cdot\textbf{1}_2$ . 
\\

For the left-handed spinor, the Dirac operator is given by
\begin{eqnarray}
  \sigma^\mu\Pt_\mu
  &=&
  \frac{\sigma^1+i\sigma^2}{2}
  \left(
    \Pt_1-i\Pt_2
  \right)
  +
  \frac{\sigma^1-i\sigma^2}{2}
  \left(
    \Pt_1+i\Pt_2
  \right)
  \nonumber\\
  & &
  \quad
  +
  \frac{\sigma^3+i\sigma^4}{2}
  \left(
    \Pt_3-i\Pt_4
  \right)
  +
  \frac{\sigma^3-i\sigma^4}{2}
  \left(
    \Pt_3+i\Pt_4
  \right)
  \nonumber\\
  &=&
  2\sqrt{N}
  \left(
    \begin{array}{cc}
      L_2[V,\ \cdot\ ]  &L_1[U^{-1},\ \cdot\ ]\\
      L_1[U,\ \cdot\ ]  &-L_2[V^{-1},\ \cdot\ ]
    \end{array}
  \right). 
\end{eqnarray}
Hence, there are two left-handed zero modes, 
\begin{eqnarray}
  \left(  
    \begin{array}{c}
      L_2V^{-1}\\
      -L_1U
    \end{array}
  \right)
  \quad
  {\rm and}
  \quad
  \left(  
    \begin{array}{c}
      L_1U^{-1}\\
      L_2V
    \end{array}
  \right). 
\end{eqnarray} 

Similarly for the right-handed spinors we have two zero modes given by
\begin{eqnarray}
  \left(  
    \begin{array}{c}
      L_2V\\
      -L_1U
    \end{array}
  \right)
  \quad
  {\rm and}
  \quad
  \left(  
    \begin{array}{c}
      L_1U^{-1}\\
      L_2V^{-1}
    \end{array}
  \right).
\end{eqnarray}
These zero-modes are not doublers, because their momenta become 
zero in the large-$N$ limit.

\subsection{Zero and tachyonic modes of the vector} \label{subsec:tachyon_boson}

Although the numbers of zero and tachyonic modes depend on $N$, they
become constant when $N$ is sufficiently large. 
If the ratio $r=\frac{L_2}{L_1}=1$, there are 4 zero modes and 5
tachyons. On the other hand, if $r\neq 1$, there are no zero modes and
7 tachyons. 

A direct calculation shows that if $r=1$ , 
$(P_1,P_2,-P_3,-P_4)$ is an eigenstate with the smallest eigenvalue  
$-16NL^2\sin^2\left(\frac{\pi}{N}\right)$. This mode corresponds to 
the deformation of the ratio $r$.

\section{2-loop free energy : $O(\Ft^0)$ and $O(\Ft^2)$ contributions}
\label{twoener}
\setcounter{equation}{0}

Here
we evaluate the $O(\Ft^0)$ and  $O(\Ft^2)$ contributions to the  2-loop free energy 
on the fuzzy $T^4$. 
For brevity of notation, we introduce the following notations
\begin{eqnarray}
  F_1(m)
  \equiv
  \frac{1}{A}\sin^2\left(\frac{\pi m_1}{N}\right)\ ,~~ 
  && 
  F_2(m)
  \equiv
  -
  \frac{1}{A               }\sin\left(\frac{\pi m_1}{N}\right)
  \cos\left(\frac{\pi m_1}{N}\right),
  \nonumber\\
  F_3(m)
  \equiv
  -\frac{1}{A}\sin^2\left(\frac{\pi m_2}{N}\right),
  && 
  F_4(m)
  \equiv  
  \frac{1}{A}\sin\left(\frac{\pi m_2}{N}\right)
  \cos\left(\frac{\pi m_2}{N}\right),
  \nonumber\\
  F_5(m)
  \equiv
  A\sin^2\left(\frac{\pi m_3}{N}\right),~~ 
  && 
  F_6(m)
  \equiv
  -
  A\sin\left(\frac{\pi m_3}{N}\right)
  \cos\left(\frac{\pi m_3}{N}\right),
  \nonumber\\
  F_7(m)
  \equiv
  -A\sin^2\left(\frac{\pi m_4}{N}\right),
  && 
  F_8(m)
  \equiv  
  A\sin\left(\frac{\pi m_4}{N}\right)
  \cos\left(\frac{\pi m_4}{N}\right),
\nonumber
\end{eqnarray}
where $A=\sqrt{r\sqrt{R}}$. We further introduce the abbreviation $F(m)\cdot
F(n)=\sum_j F_j(m)F_j(n)$, which is the analogue of the inner product
of momenta.

\subsection{$O(\Ft^0)$}

We have four kinds of 1PI 2-loop vacuum diagrams ( Fig
\ref{fig:2loop_diagrams}). We list the contributions from each diagram
: 
\\
\\
\textbf{(a) Diagram with two fermion-fermion-boson vertices $V_{3f}$} 
\begin{eqnarray}
  \lefteqn{
    \frac{g^2}{4N^2L^4}
    \sum_{m,n}
    \frac{F(m)\cdot F(-n)}{F(m)^2\cdot F(n)^2\cdot F(m+n)^2}
    \left(1-\cos(2\pi m\circ n)\right)
  }\nonumber\\
  &=&
  \frac{g^2}{8N^2L^4}
  \sum_{m,n}
  \frac{1}{F(m)^2\cdot F(n)^2}
  \left(1-\cos(2\pi m\circ n)\right). 
  \label{eq:2loop_V3f}
\end{eqnarray}
Here we have used 
$2F(m)\cdot F(-n)=F(m)^2+F(n)^2+F(m+n)^2$. 
\\
\\
\textbf{(b) Diagram with two ghost-ghost-vector vertices $V_{3g}$} 
\\

The exact result is given by
\begin{eqnarray}
  \lefteqn{
    -\frac{g^2}{512N^2L^4}
    \sum_{\pm}\sum_{m,n}
    \biggl\{
    \frac{L_1^2\sin\left(\frac{\pi m_2}{N_1}\right)
        \sin\left(\frac{\pi n_2}{N_1}\right)}
      {F(m_1+n_1\pm 1,m_2+n_2,m_3+n_3,m_4+n_4)^2\cdot F(m)^2\cdot F(n)^2}
    }
    \nonumber\\
    & &
    \times  
    \left\{
      \cos\left(2\pi m\circ n\pm\pi\frac{n_2-m_2}{N_1}\right)
      -
      \cos\left(\pi\frac{m_2+n_2}{N_1}\right)
    \right\}
    \nonumber\\
    & &
    +
    \frac{L_1^2\sin\left(\frac{\pi m_1}{N_1}\right)
      \sin\left(\frac{\pi n_1}{N_1}\right)}
    {F(m_1+n_1,m_2+n_2\pm 1,m_3+n_3,m_4+n_4)^2\cdot F(m)^2\cdot F(n)^2}
  \nonumber\\
  & &
  \times  
  \left\{
    \cos\left(2\pi m\circ n\mp\pi\frac{n_1-m_1}{N_1}\right)
    -
    \cos\left(\pi\frac{m_1+n_1}{N_1}\right)
  \right\}
  \biggl\}
  \nonumber\\
  & &
  +
  \left(
    L_1,N_1,m_1,m_2,n_1,n_2
    \leftrightarrow
    L_2,N_2,m_3,m_4,n_3,n_4
  \right). 
  \label{eq:2loop_V3g_exact}
\end{eqnarray}
Shifting $m_i$'s and introducing approximations,
\begin{eqnarray}
  \frac{1}{F(m_1+n_1\pm 1,m_2+n_2,m_3+n_3,m_4+n_4)^2}
  \simeq
  \frac{1}{F(m_1+n_1,m_2+n_2,m_3+n_3,m_4+n_4)^2} 
  \label{simeq:approximation}
\end{eqnarray}
and so on, we obtain 
\begin{eqnarray}
  -\frac{g^2}{512N^2L^4}
  \sum_{m,n}
  \frac{1-\cos(2\pi m\circ n)}{F(m)^2\cdot F(n)^2}.
  \label{eq:2loop_V3g_approximated}
\end{eqnarray}
\\
\textbf{(c) Diagram with two boson-boson-vector vertices $V_{3b}$} 
\\

The exact result is given by 
\begin{eqnarray}
   \lefteqn{
    \frac{9g^2L_1^2}{512N^2}
    \sum_{\pm}\sum_{m,n}
    \frac{1}{F(m_1+n_1\mp 1,m_2+n_2,m_3+n_3,m_4+n_4)^2\cdot F(m)^2\cdot F(n)^2}
  }\nonumber\\
  & &
  \times
  \biggl\{
  \sin^2\left(\frac{\pi m_2}{N_1}\right)
  \left\{
    1-\cos\left(2\pi m\circ n\mp\frac{2\pi n_2}{N_1}\right)
  \right\}  
  \nonumber\\
  & &
  \qquad
  -
  \sin\left(\frac{\pi m_2}{N_1}\right)
  \sin\left(\frac{\pi n_2}{N_1}\right)
  \left\{
    \cos\left(\frac{\pi (m_2+n_2)}{N_1}\right)
    -
    \cos\left(2\pi m\circ n\pm\frac{\pi(m_2-n_2)}{N_1}\right)
  \right\}
  \biggl\}
  \nonumber\\
  \lefteqn{
    +
    \frac{9g^2L_1^2}{512N^2}
    \sum_{\pm}\sum_{m,n}
    \frac{1}{F(m_1+n_1,m_2+n_2\pm 1,m_3+n_3,m_4+n_4)^2\cdot F(m)^2\cdot F(n)^2}
  }\nonumber\\
  & &
  \times
  \biggl\{
  \sin^2\left(\frac{\pi m_1}{N_1}\right)
  \left\{
    1-\cos\left(2\pi m\circ n\mp\frac{2\pi n_1}{N_1}\right)
  \right\}  
  \nonumber\\
  & &
  \qquad
  -
  \sin\left(\frac{\pi m_1}{N_1}\right)
  \sin\left(\frac{\pi n_1}{N_1}\right)
  \left\{
    \cos\left(\frac{\pi (m_1+n_1)}{N_1}\right)
    -
    \cos\left(2\pi m\circ n\pm\frac{\pi(m_1-n_1)}{N_1}\right)
  \right\}
  \biggl\}
  \nonumber\\
  & &
  +
  \left(
    L_1,N_1,m_1,m_2,n_1,n_2
    \leftrightarrow
    L_2,N_2,m_3,m_4,n_3,n_4
  \right).
  \label{eq:2loop_V3b_exact}
\end{eqnarray}
Again by shifting $m$ and using the approximations (\ref{simeq:approximation}), we can rewrite it as 
\begin{eqnarray}
  \frac{27g^2}{512N^2}
  \sum_{m,n}\frac{1-\cos(2\pi m\circ n)}{F(m)^2\cdot F(n)^2}. 
  \label{eq:2loop_V3b_approximated}
\end{eqnarray}
\\
\textbf{(d) Diagram with two 4-boson vertex $V_{4b}$} 
\begin{eqnarray}
  -\frac{45g^2}{256N^2}
  \sum_{m,n}\frac{1-\cos(2\pi m\circ n)}{F(m)^2\cdot F(n)^2}. 
  \label{eq:2loop_V4b}
\end{eqnarray}

Summing up (a)$\sim$(d), we find that the 2-loop vacuum diagrams almost cancel. Actually, if we use the approximate values (\ref{eq:2loop_V3g_approximated}) and (\ref{eq:2loop_V3b_approximated}), we have 
\begin{eqnarray}
  (\ref{eq:2loop_V3f})
  +
  (\ref{eq:2loop_V3g_approximated}) 
  +
  (\ref{eq:2loop_V3b_approximated})
  +
  (\ref{eq:2loop_V4b})
  =
  \left(
    \frac{1}{8}
    -
    \frac{1}{512}
    +
    \frac{27}{512}
    -
    \frac{45}{256}
  \right)
  \frac{g^2}{N^2}
  \sum_{m,n}\frac{1-\cos(2\pi m\circ n)}{F(m)^2\cdot F(n)^2}
  =
  0. 
\end{eqnarray}

This cancellation is not exact, but we can see numerically that
contribution of the $O(\Ft^0)$ terms is smaller than that of the
$O(\Ft^2)$ terms. 

\subsection{$O(\Ft^2)$}

The calculation of the $O(\Ft^2)$ contribution is very tedious. Here
we show only the simplest case, 
that is, the planar part of Fig.\ref{fig:2loop_diagrams} (b) : 
\begin{eqnarray}
  \lefteqn{
    \frac{g^2\pi^2}{256L^4N^3}\sum_{m,n}
    \biggl\{
    \frac{
      1
      -
      \cos\left(\frac{2\pi(m_1+n_1)}{N_1}\right)
      \cos\left(\frac{2\pi(m_2+n_2)}{N_1}\right)
    }{F(m)^2\cdot F(n)^2\cdot(F(m+n)^2)^3}
  }\nonumber\\
  & &
  \times
  \frac{1}{r^3\sqrt{R}} 
  \left(
    \smone\snone\cmnone
    +
    \smtwo\sntwo\cmntwo
  \right)
  \nonumber\\
  & &
  +
  \frac{
      1
      -
      \cos\left(\frac{2\pi(m_3+n_3)}{N_2}\right)
      \cos\left(\frac{2\pi(m_4+n_4)}{N_2}\right)
    }{F(m)^2\cdot F(n)^2\cdot(F(m+n)^2)^3}
    \nonumber\\
  & &
  \times
  r^3\sqrt{R}
  \left(
    \smthree\snthree\cmnthree
    +
    \smfour\snfour\cmnfour
  \right) 
  \biggl\}
  \nonumber\\
  &\sim&
  \frac{g^2\pi^2N}{256L^4}\int d^4k d^4l
  \biggl\{
  \frac{
    1
    -
    \cos\left(2\pi(k_1+l_1)\right)
    \cos\left(2\pi(k_2+l_2)\right)
  }{F(k)^2\cdot F(l)^2\cdot(F(k+l)^2)^3}
  \nonumber\\
  & &
  \times
  \frac{1}{r^3\sqrt{R}} 
  \left(
    \sin(\pi k_1)\sin(\pi l_1)\cos\left(\pi(k_1+l_1)\right)
    +
    \sin(\pi k_2)\sin(\pi l_2)\cos\left(\pi(k_2+l_2)\right)
  \right)
  \nonumber\\
  & &
  +
  \frac{
      1
      -
      \cos\left(2\pi(k_3+l_3)\right)
      \cos\left(2\pi(k_4+l_4)\right)
    }{F(k)^2\cdot F(l)^2\cdot(F(k+l)^2)^3}
    \nonumber\\
  & &
  \times
  r^3\sqrt{R}
  \left(
    \sin(\pi k_3)\sin(\pi l_3)\cos\left(\pi(k_3+l_3)\right)
    +
    \sin(\pi k_4)\sin(\pi l_4)\cos\left(\pi(k_4+l_4)\right)
  \right) 
  \biggl\}. 
  \nonumber\\
\end{eqnarray} 
Here, $k_i=m_i/N_1,k_j=m_j/N_2,l_i=n_i/N_1,l_j=n_j/N_2\
(i=1,2;j=3,4)$, and $k_\mu$ and $l_\mu$ vary from $0$ to $1$.

\end{document}